\documentclass[aps,prl,english,reprint,preprintnumbers,superscriptaddress,amsmath,amssymb,bibnotes,longbibliography]{revtex4-2}
\usepackage{graphicx}
\usepackage{dcolumn}
\usepackage{epstopdf}
\usepackage{bm}
\usepackage{flushend}
\usepackage[pdfstartview=FitH, CJKbookmarks=true, bookmarksnumbered=true, bookmarksopen=true, colorlinks, pdfborder=001, linkcolor=blue, anchorcolor=blue, citecolor=blue,urlcolor=blue,breaklinks]{hyperref}
\usepackage[mathlines]{lineno}

\begin{document}
	
\title{Phonon softening and slowing down of charge-density-wave fluctuations in BaNi$_2$As$_2$}

\author{Yu Song}
\email{yusong\_phys@zju.edu.cn}
\affiliation{Center for Correlated Matter and School of Physics, Zhejiang University, Hangzhou 310058, China}
\affiliation{Zhejiang Province Key Laboratory of Quantum Technology and Device, School of Physics, Zhejiang University, Hangzhou 310058, China}

\author{Shan Wu}
\affiliation{Physics Department, University of California, Berkeley, California 94720, USA}
\affiliation{Materials Science Division, Lawrence Berkeley National Lab, Berkeley, California 94720, USA}

\author{Xiang Chen}
\affiliation{Physics Department, University of California, Berkeley, California 94720, USA}
\affiliation{Materials Science Division, Lawrence Berkeley National Lab, Berkeley, California 94720, USA}

\author{Yu He}
\affiliation{Department of Applied Physics, Yale University, New Haven, Connecticut 06511, USA}
\affiliation{Physics Department, University of California, Berkeley, California 94720, USA}
\affiliation{Materials Science Division, Lawrence Berkeley National Lab, Berkeley, California 94720, USA}

\author{Hiroshi Uchiyama}
\affiliation{Japan Synchrotron Radiation Research Institute, SPring-8, 1-1-1 Kouto, Sayo, Hyogo 679-5198, Japan}

\author{Baizhuo Li}
\affiliation{Zhejiang Province Key Laboratory of Quantum Technology and Device, School of Physics, Zhejiang University, Hangzhou 310058, China}

\author{Saizheng Cao}
\affiliation{Center for Correlated Matter and School of Physics, Zhejiang University, Hangzhou 310058, China}

\author{Jiayu Guo}
\affiliation{Center for Correlated Matter and School of Physics, Zhejiang University, Hangzhou 310058, China}

\author{Guanghan Cao}
\affiliation{Zhejiang Province Key Laboratory of Quantum Technology and Device, School of Physics, Zhejiang University, Hangzhou 310058, China}

\author{Robert Birgeneau}
\affiliation{Physics Department, University of California, Berkeley, California 94720, USA}
\affiliation{Materials Science Division, Lawrence Berkeley National Lab, Berkeley, California 94720, USA}

\begin{abstract}
BaNi$_2$As$_2$ is a nonmagnetic analogue of the iron pnictide superconductors, and in its tetragonal state exhibits an incommensurate charge-density-wave (IC-CDW) and a sizable elastoresistance.
In this work, phonons in BaNi$_2$As$_2$ associated with the IC-CDW and uniform in-plane lattice distortions are investigated using high-resolution inelastic X-ray scattering. The in-plane transverse acoustic phonons reveal no softening at temperatures where the elastoresistance increases strongly, indicating the latter to be electronically driven. Systematic phonon measurements reveal the IC-CDW occurs in two stages upon cooling: underdamped phonons first soften to zero energy well above the IC-CDW ordering temperature, then the resulting quasielastic IC-CDW fluctuations gradually slow down and coalesce into the static IC-CDW order. 
A possible origin for our observations is the IC-CDW in tetragonal BaNi$_2$As$_2$ being uniaxial, which provides an additional Ising degree of freedom favorable for disordered IC-CDW modulations, and accounts for the elastoresistance through a weak coupling to the lattice.
\end{abstract}

\maketitle

How charge-density-waves (CDWs) arise is a persistent question in condensed matter physics \cite{Zhu2015,Zhu2017,Chen2016}, as it sheds light on the microscopic origins for the CDW instability and how CDWs interact with orders such as superconductivity \cite{Gabovich2001,Comin2016,Frano2020} and electronic nematicity \cite{Nie2014,Nie2022,Grissonnanche2022}. In Peierls' model, CDW order in an one-dimensional system emerges when Fermi surface nesting drives a sharp Kohn anomaly towards zero energy \cite{Zhu2017}. In the archetypal quasi-two-dimensional CDW material $2H$-NbSe$_2$, phonon softening driven by electron-phonon coupling occurs over an extended region in reciprocal space and the CDW order sets in when phonons soften to zero energy \cite{Weber2011}. In the cuprates, a partial softening of phonons leads to a two-dimensional CDW \cite{LeTacon2013,YuHe2018}, whereas a three-dimensional CDW is associated with phonons that soften to zero energy \cite{Kim2018}. A common theme of these CDW transitions is that they are dominantly driven by phonon softening, whereas studies of CDW transitions that occur through the slowing down of quasielastic fluctuations have been limited \cite{Torchinsky2013,Zong2019,Upreti2022}.

Recently, a variety of CDW states were uncovered in Ba(Ni$_{1-x}$Co$_x$)$_2$As$_2$ and Ba$_{1-x}$Sr$_x$Ni$_2$As$_2$ \cite{Lee2019,Lee2021}. In combination with a superconducting ground state that becomes enhanced due to strong nematic fluctuations \cite{Eckberg2019}, these materials are excellent candidates for investigating how CDW states emerge and interact with nematicity and superconductivity \cite{Eckberg2019,Lederer2020}. BaNi$_2$As$_2$, the parent compound of these materials, is a nonmagnetic analogue of the 122-family of iron pnictides, and becomes superconducting below $T_{\rm c}\approx0.7$~K \cite{Ronning2008,Sefat2009}. Upon cooling from high temperatures, BaNi$_2$As$_2$ first enters into an incommensurate CDW (IC-CDW) with a propagation vector ${\bf q}_0\approx(0.28,0,0)/(0,0.28,0)$ below $T_{\rm IC}\approx150$~K, then exhibits a strongly first-order structural transition at $T_{\rm S}\approx136$~K and transitions into a triclinic phase, which hosts a commensurate CDW (C-CDW) \cite{Lee2019}. 
While the C-CDW in triclinic BaNi$_2$As$_2$ is found to be uniaxial \cite{Lei2022,Guo2022}, the IC-CDW in BaNi$_2$As$_2$ has been suggested to be either uniaxial \cite{Lee2019} or biaxial \cite{Lee2021}, and in the uniaxial case it may be associated with orbital ordering \cite{Merz2021,Pokharel2021}. No energy gap or band folding is observed in photoemission measurements for the IC-CDW \cite{Zhou2011,Guo2022}, which suggests that it may be unconventional \cite{Lee2019}. 

In the iron pnictides, the coupling between nematicity and the lattice leads to a divergent elastoresistance in the $B_{2g}$ channel \cite{Chu2012} and a tetragonal-to-orthorhombic structural transition driven by the softening of in-plane transverse acoustic (IPTA) phonons \cite{Niedziela2011,Parshall2015,Li2018,Weber2018,Merritt2020,Kauth2020,Wu2021}. 
In BaNi$_2$As$_2$, a sizable elastoresistance in the $B_{\rm 1g}$ channel appears upon cooling towards $T_{\rm IC}$, and an orthorhombic distortion of the lattice is reported for the IC-CDW phase \cite{Merz2021}. The elastoresistance becomes further enhanced and hysteretic inside the IC-CDW state, suggesting the IC-CDW state being associated with the ordering of a nematic order parameter \cite{Eckberg2019}. Whether the enhancement of elastoresistance in BaNi$_2$As$_2$ is associated with softening of the IPTA phonons, and how the elastoresistance may be related to the IC-CDW, remain unresolved.

\begin{figure}
	\includegraphics[width=1\columnwidth]{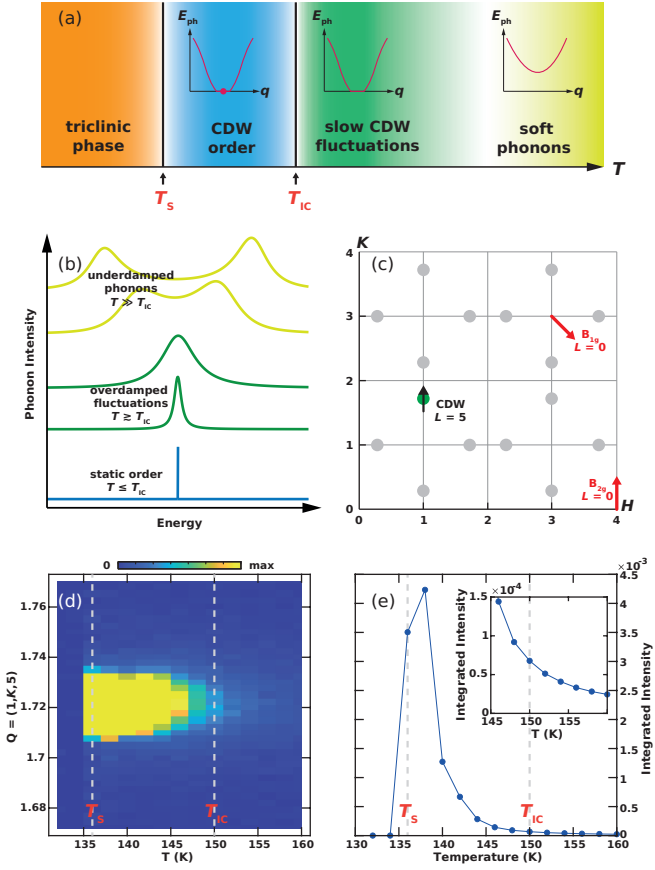} \protect\caption{
	(a) Schematic evolution of BaNi$_2$As$_2$ with temperature. The IC-CDW order forms at $T_{\rm IC}$, and a structural transition occurs at $T_{\rm S}$.
	The insets in the three regions for $T>T_{\rm S}$ are schematic phonon dispersions around ${\bf q}_0$. (b) Schematic energy scans in BaNi$_2$As$_2$ at ${\bf q}_0$ for $T>T_{\rm S}$.
	(c) Schematic of the reciprocal space probed in this work. The light gray circles are IC-CDW peaks which occur for odd $L$, and the black arrow represents a line along $(1,K,5)$, where our measurements are carried out. The red arrows present measurements of the IPTA phonons, with lattice distortions in the $B_{\rm 1g}$ and $B_{\rm 2g}$ channels respectively probed by phonons at $(3+q,3-q,0)$ and $(4,k,0)$. (d) Color-coded intensity of elastic scans along $(1,K,5)$ as a function of temperature. Alignment in the tetragonal phase is used for scans in the triclinic phase. (e) Temperature dependence of the integrated intensity for scans in (d). The inset zooms in around 150~K.
	}
	\label{Fig1}
\end{figure}

In this work, we use high-resolution inelastic X-ray scattering to show that phonons in BaNi$_2$As$_2$ soften in an extended region around the IC-CDW ordering vector, and upon cooling become overdamped prior to the appearance of static IC-CDW order [Figs.~\ref{Fig1}(a) and (b)]. These overdamped phonons correspond to fluctuations that relax in time, with the relaxation rate at the ordering vector linearly extrapolating to zero around $T_{\rm IC}$, indicating the slowing down of IC-CDW fluctuations. By measuring the dispersion of IPTA phonons in BaNi$_2$As$_2$, we find no detectable softening associated with nematicity as in the iron pnictides, pointing to a weak coupling between nematicity and the lattice. This in turns suggests that the sizable elastoresistance is BaNi$_2$As$_2$ is electronically driven and closely associated with the growth of IC-CDW correlations, whereas the lattice plays a minor role.

BaNi$_2$As$_2$ single crystals were synthesized using a Pb-flux method \cite{Ronning2008}. Inelastic X-ray scattering measurements with the photon energy at 21.7476~keV and an energy resolution of $\sim1.4$~meV were carried out using the BL35XU beamline \cite{Baron2000} at SPring-8, Japan. Momentum transfer ${\bf Q}$ are referenced in reduced lattice units, using the tetragonal cell of BaNi$_2$As$_2$ with lattice constants $a=b\approx4.13$~{\AA} and $c\approx11.6$~{\AA}. Measured phonon spectra 
are fit to a general damped harmonic oscillator (DHO) \cite{Lamsal2016,Robarts2019} or a Lorentzian function, after convolution with the instrumental resolution. Additional details on the experiment and data analysis are provided in the Supplemental Materials \cite{SM}.


The IC-CDW peaks in BaNi$_2$As$_2$ exhibit fourfold rotational symmetry [light gray circles in Fig.~\ref{Fig1}(c)], which may be intrinsic due to a biaxial CDW \cite{Lee2021} or result from the twinning of a uniaxial CDW \cite{Lee2019,Merz2021}. To probe the temperature evolution of the IC-CDW, scans along ${\bf Q}=(1,K,5)$ are carried out around $K=1.72$ [Fig.~\ref{Fig1}(c)] upon cooling across $T_{\rm IC}$, revealing the IC-CDW propagating vector ${\bf q}_0$ to be independent of temperature. From the color-coded plot [Fig.~\ref{Fig1}(d)] and temperature dependence of the integrated peak intensity [Fig.~\ref{Fig1}(e)], we find that the IC-CDW peak onsets around $T_{\rm IC}$, experiences a rather gradual increase upon cooling, and becomes completely quenched upon entering the triclinic phase below $T_{\rm S}$, consistent with previous report \cite{Lee2019}. 

\begin{figure*}
	\includegraphics[width=\linewidth]{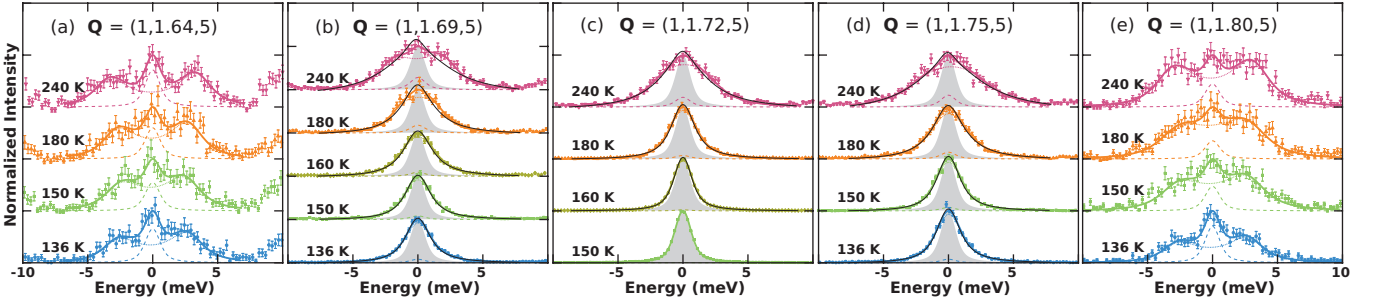} \protect\caption{
		Phonon measurements at several temperatures for (a) ${\bf Q}=(1,1.64,5)$, (b) ${\bf Q}=(1,1.69,5)$, (c) ${\bf Q}=(1,1.72,5)$, (d) ${\bf Q}=(1,1.75,5)$, and (e) ${\bf Q}=(1,1.80,5)$. The solid lines are fits to either the DHO model (thick lines with same color as the symbols) or a Lorentzian model (thin black lines). 
		The dashed lines are the elastic components in the DHO model, and the dotted lines are the inelastic components. The experimentally measured instrumental resolution is parametrized using a pseudo-Voigt function, and is represented as the shaded gray areas.
	}
	\label{Fig2}
\end{figure*}

To probe how the IC-CDW emerges in BaNi$_2$As$_2$, phonons at ${\bf Q}=(1,K,5)$ with $K=1.64$, 1.69, 1.72, 1.75 and 1.80 were measured at several temperatures both above and below $T_{\rm IC}$, shown in Fig.~\ref{Fig2}. For $K=1.64$ and 1.80 [Figs.~\ref{Fig2}(a) and (e)], the phonons are clearly underdamped with well-resolved energies, and do not change significantly with temperature. On the other hand, for $K=1.69$, 1.72 and 1.75 [Figs.~\ref{Fig2}(b)-(d)], the phonons form a single peak centered around zero energy and sharpen upon cooling. For $K=1.72$, the signal becomes resolution-limited at 150~K [shaded gray area], consistent with the onset of IC-CDW order at $T_{\rm IC}$.

To quantitatively analyze the dynamics associated with the IC-CDW in BaNi$_2$As$_2$, we fit the data in Fig.~\ref{Fig2} to a DHO function \cite{Lamsal2016,Robarts2019,SM}:
\begin{equation}
S(E)=\frac{A}{1-\exp(-\frac{E}{k_{\rm B}T})}\frac{\gamma E}{(E^2-E_0^2)^2+(E\gamma)^2},
\label{DHO}
\end{equation}
shown as thick solid lines. In the DHO model, $A$ is an intensity scale factor, $E_0$ is the undamped phonon energy, and $\gamma$ is the damping rate.
A general DHO is underdamped when $\gamma/(2E_0)<1$, and in this case the DHO describes phonons with an energy $E_{\rm ph}=\sqrt{E_0^2-\gamma^2/4}$. When $\gamma/(2E_0)>1$, the DHO describes overdamped quasielastic excitations that do not oscillate but only decay in time. 

\begin{figure}
	\includegraphics[width=1\columnwidth]{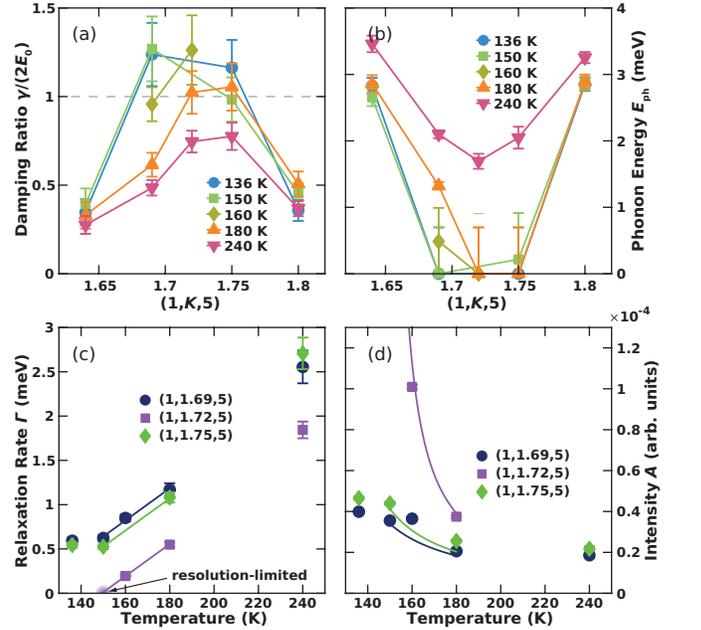} \protect\caption{
	(a) Damping ratios $\gamma/(2E_0)$, and (b) phonon energies $E_{\rm ph}$, obtained by fitting the data in Fig.~\ref{Fig2} to a general DHO. (c) Relaxation rates $\Gamma$, and (d) the intensity scale factors $A$, obtained by fitting the data in Figs.~\ref{Fig2}(b)-(d) to a Lorentzian model. The solid lines in (c) are a fit to Eq.~\ref{Gamma}, 
	for data inclusively between 150~K and 180~K. The solid lines in (d) are a fit to Eq.~\ref{A},
	with $A_0$ as the only free parameter and the other parameters are from the fit in (c). 
 	}
	\label{Fig3}
\end{figure}

The fit values of $\gamma/(2E_0)$ are shown in Fig.~\ref{Fig3}(a), revealing that while the phonons along $(1,K,5)$ are underdamped at 240~K, some already become overdamped at 180~K, well above $T_{\rm IC}\approx150$~K. The corresponding phonon energies $E_{\rm ph}$ [Fig.~\ref{Fig3}(b)] exhibit clear softening over an extended region at 240~K, similar to $2H$-NbSe$_2$ \cite{Weber2011}. This suggests a dominant role of electron-phonon coupling in the formation of the IC-CDW in BaNi$_2$As$_2$, consistent with the absence of Fermi surface nesting associated with the IC-CDW ordering vector in photoemission measurements \cite{Zhou2011,Guo2022}.
On the other hand, whereas the softening of phonons to zero energy occurs well above $T_{\rm IC}$ in BaNi$_2$As$_2$, it happens only right around the CDW ordering temperature in $2H$-NbSe$_2$ \cite{Weber2011}. 

Motivated by the observation of overdamped phonons well above $T_{\rm IC}$, the dynamics at ${\bf Q}=(1,K,5)$ with $K=1.69$, 1.72, and 1.75 are fit to a Lorentzian model:
\begin{equation}
S(E)=\frac{A}{1-\exp(-\frac{E}{k_{\rm B}T})}\frac{\Gamma E}{E^2+\Gamma^2},
\label{Lorentz}
\end{equation}   
shown as thin black lines in Figs.~\ref{Fig2}(b)-(d) \cite{SM}. In the Lorentzian model, $A$ is an intensity scale factor, and $\Gamma$ is the relaxation rate. At 240~K, the Lorentzian model clearly deviate from the data, consistent with the underdamped nature of the corresponding DHO fits. At 180~K and lower temperatures, much better agreements are found, consistent with the phonons at these temperatures being overdamped or close to overdamped. From fits to the Lorentzian model, we extract the values of $\Gamma$ and $A$, which are respectively shown in Figs.~\ref{Fig3}(c) and (d).


From the relaxation rate $\Gamma$ for the three wave vectors in Fig.~\ref{Fig3}(c), we observe that (1) $\Gamma$ appears linear in temperature for $T\gtrsim T_{\rm IC}$, and the slopes for such a linear behavior does not change significantly between the three wave vectors; (2) for $K=1.69$ and 1.75, which are equally distant from $K=1.72$, their $\Gamma$ lines are almost equally displaced from the $K=1.72$ line; (3) $\Gamma$ for $K=1.72$ extrapolates to zero at around 150~K, consistent with the experimental data at this temperature being resolution-limited [Fig.~\ref{Fig2}(c)]. Interestingly, such behaviors for $\Gamma$ are in line with the model of magnetic fluctuations in nearly antiferromagnetic metals \cite{Moriya}, which was successfully used to describe the quasielastic spin dynamics in the iron pnictides \cite{Inosov2009}. A similar model was also used to describe the critical orientation dynamics in the plastic crystal CD$_4$ \cite{Huller1972,Press1974}.

In the model for nearly antiferromagnetic metals, the relaxation rate in the disordered state behaves as:
\begin{equation}
\Gamma=\Gamma_0[T+\theta+\xi_0^2 ({\bf q}-{\bf q}_0)^2],
\label{Gamma}
\end{equation}
where $\Gamma_0$ and $\xi_0$ are respectively related to the temperature dependence of the relaxation rate and the correlation length, ${\bf q}_0$ is the ordering vector and $\theta$ is the Curie-Weiss temperature \cite{Inosov2009}. The data for the three wave vectors at 150~K, 160~K and 180~K [Fig.~\ref{Fig3}(c)] are collectively fit to Eq.~\ref{Gamma} (shown as solid lines), resulting in an excellent description of the data with ${\bf q}_0=(0,0.2785(4),0)$, $\theta=-149(1)$~K, $\Gamma_0=0.018(1)$~meV$\cdot$K$^{-1}$, and $\xi_0=123(4)$~{\AA}$\cdot$K$^{-1/2}$. The same model predicts that the intensity factor $A$ of the Lorentzian function behaves as: 
\begin{equation}
A=\frac{A_0}{T+\theta+\xi_0^2 ({\bf q}-{\bf q}_0)^2},
\label{A}
\end{equation}
by using the parameters from the fit in Fig.~\ref{Fig3}(c) and introducing an additional intensity scale factor $A_0$, we find the model also qualitatively captures how the intensity factor $A$ evolves with temperature and wave vector [solid lines in Fig.~\ref{Fig3}(d)]. At 136~K ($T<T_{\rm IC}$), $\Gamma$ for $K=1.69$ and 1.75 clearly deviate from the linear trend for $T\gtrsim T_{\rm IC}$, consistent with the appearance of static IC-CDW order below $T_{\rm IC}$.    

\begin{figure}
	\includegraphics[width=1\columnwidth]{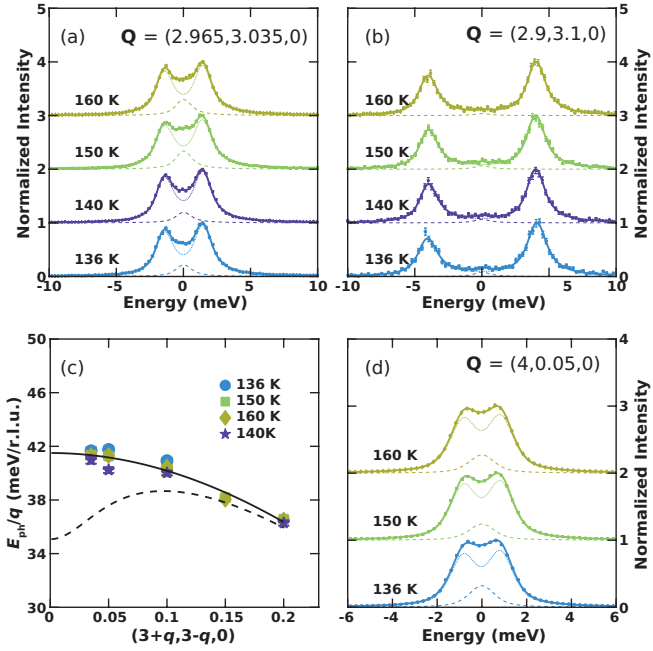} \protect\caption{
		Temperature evolution of acoustic phonons at ${\bf Q}=(3+q,3-q,0)$ with (a) $q=0.035$ and (b) $q=0.1$. (c) $E_{\rm ph}/q$ for phonons at ${\bf Q}=(3+q,3-q,0)$, the solid line is a fit to $E_{\rm ph}/q=B\sin(\pi D q)$, and the dashed line schematically shows the expected behavior for $E_{\rm ph}/q$ when the lattice couples strongly to nematic fluctuations. (d) Temperature evolution of acoustic phonons at ${\bf Q}=(4,k,0)$ for $k=0.05$. The solid lines in (a), (b) and (d) are fits to the DHO model.
	}
	\label{Fig4}
\end{figure}

Having established that the IC-CDW in BaNi$_2$As$_2$ emerges through a slowing down process, the evolution of IPTA phonons are probed around $T_{\rm IC}$ upon cooling, as the IC-CDW correlations build up and the elastoresistance grows. Measurements were carried out at ${\bf Q}=(3+q,3-q,0)$ and ${\bf Q}=(4,k,0)$, which are dominated by IPTA phonons [Fig.~\ref{Fig1}(c)]. Due to having the same symmetry, these IPTA phonons directly couple to nematic order parameters in the $B_{\rm 1g}$ and $B_{\rm 2g}$ channels, respectively. Figs.~\ref{Fig4}(a) and (b) show the evolution of ${\bf Q}=(3+q,3-q,0)$ IPTA phonons with temperature for $q=0.035$ and $0.1$, and in both cases we find the IPTA phonons do not exhibit a clear change in energy between 160~K to 136~K, in strong contrast to the elastoresistance which grows substantially in the same temperature range \cite{Eckberg2019}. The dispersion of $E_{\rm ph}({\bf Q})$ around ${\bf Q}=(3+q,3-q,0)$ is extracted by fitting the data to the general DHO model [solid lines in Figs.~\ref{Fig4}(a) and (b)], and to accentuate the behavior of phonons at small $q$, $E_{\rm ph}/q$ is shown in Fig.~\ref{Fig4}(c).

The essentially temperature-independent dispersion of IPTA phonons in BaNi$_2$As$_2$ can be described by $E_{\rm ph}=B\sin(D\pi q)$ [solid line in Fig.~\ref{Fig4}(c)] without invoking the coupling to another order parameter. This is different from the iron pnictides, where due to coupling with nematicity, clear softening of the IPTA phonons at small momenta are observerd \cite{Niedziela2011,Parshall2015,Li2018,Weber2018,Merritt2020,Kauth2020,Wu2021}. This coupling would manifest as a substantial drop in the value of $E_{\rm ph}/q$ at small $q$ [dashed line in Fig.~\ref{Fig4}(c)], clearly absent in our data.

Our fit of the phonon dispersion at $(3+q,3-q,0)$ gives a sound velocity of $19.3(1)$~meV$\cdot${\AA} (2.9~km/s), and the associated shear modulus is $(C_{11}-C_{12})/2=58.4(3)$~GPa. For completeness, the IPTA phonons at ${\bf Q}=(4,0.05,0)$ were also measured [Fig.~\ref{Fig4}(d)], and are also found to be independent of temperature across $T_{\rm IC}$. From these measurements, we estimate the sound velocity to be $11.0(1)$~meV$\cdot${\AA} (1.7~km/s), and the shear modulus to be $C_{66}=19.1(2)$~GPa. The observation of $(C_{11}-C_{12})/2>C_{66}$ is consistent with the absence of lattice softening in the $B_{\rm 1g}$ channel. In the iron pnictides, a significant lattice softening \cite{Niedziela2011,Parshall2015,Li2018,Weber2018,Merritt2020,Kauth2020,Wu2021} accompanies the growth of elastoresistance, and through a divergence of the latter the electronic origin of the nematic order parameter was identified \cite{Chu2012}. In BaNi$_2$As$_2$, since the lattice remains stiff [Fig.~\ref{Fig4}(c)] as the elastoresistance grows, this directly indicates that the nematicity in BaNi$_2$As$_2$ is not lattice-driven.
  
Given BaNi$_2$As$_2$ is nonmagnetic, a candidate for its elastoresistance and nematicity is the IC-CDW being uniaxial: strain acts to detwin the sample for $T<T_{\rm IC}$ and favors one type of uniaxial IC-CDW fluctuations for $T>T_{\rm IC}$, both leading to anisotropy in electrical transport. This is reminiscent of what happens in the iron pnictides, where strain favorably selects one type of magnetic order \cite{Dhital2012,Song2013} or magnetic 
fluctuations \cite{Lu2014,Song2015}. In the scenario of the IC-CDW being uniaxial, the coherence of an Ising degree of freedom [whether the ordering vector is $(0,0.28,0)$ or $(0.28,0,0)$] is required for static order to onset, in addition to phase coherence of the IC-CDW modulation.
This is favorable for a disordered regime of IC-CDW modulations above $T_{\rm IC}$, and for the static IC-CDW order to emerge through a slowing down process.

Our findings of soft phonons and overdamped quasielastic excitations in BaNi$_2$As$_2$ suggest that the IC-CDW forms in two stages [Fig.~\ref{Fig1}(a)]. Firstly, the underdamped phonons around ${\bf q}_0$ soften to zero energy and become overdamped (occurring between 240~K and 180~K), with the IC-CDW modulations fluctuating in time. Then upon further cooling, the IC-CDW fluctuations slow down and become static at $T_{\rm IC}$ [Fig.~\ref{Fig1}(b)]. 
Our observation of quasielastic IC-CDW fluctuations in BaNi$_2$As$_2$ is reminiscent of order-disorder structural transitions \cite{Bruce1980} and superconductivity limited by phase coherence \cite{YuHe2021}, where amplitude of the order parameter forms prior to phase coherence. Continuous magnetic transitions typically arise through a slowing down process with similar quasielastic fluctuations, as in these cases the order parameter amplitude corresponds to spins, which persist above the ordering temperature. In contrast, most CDW orders form when phonons soften to zero energy, analogous to displacive structural transitions and superconductivity limited by pairing strength.

In conclusion, we find phonons around the IC-CDW ordering vector in BaNi$_2$As$_2$ soften over an extended region in momentum space, and the static IC-CDW order does not immediately form then phonons soften to zero energy. Instead, the IC-CDW dynamics exhibit a slowing down regime before coalescing into static order. The in-plane transverse acoustic phonons associated with $B_{1g}$ distortions hardly change around the IC-CDW ordering temperature, suggesting the enhancement of elastoresistance is not lattice-driven, but likely associated with the growth of IC-CDW correlations. 
Our results motivate studies on BaNi$_2$As$_2$ and its doped variants for the critical dynamics of CDW transitions, as well as the interplay between CDW fluctuations, nematicity and superconductivity.

We acknowledge Guowei Yang and Yang Liu for helpful discussions. The work at Zhejiang University was supported by the Pioneer and Leading Goose R\&D Program of Zhejiang (2022SDXHDX0005), and the Key R\&D Program of Zhejiang Province, China (2021C01002). The work at University of California, Berkeley and Lawrence Berkeley National Laboratory was supported by the Office of Science, Office of BES, Materials Sciences and Engineering Division, of the U.S. DOE under Contract No. DE-AC02-05-CH11231 within the Quantum Materials Program (KC2202). Measurements at the BL35XU beam line at SPring-8 under approval from JASRI were performed under Proposal No. 2021B1165. 

\bibliographystyle{apsrev4-2}
\bibliography{bibfile}
\end{document}